# Direct phase modulation via optical injection: theoretical study

**Roman Shakhovoy[1,2*], Marius Puplauskis[3,4], Violetta Sharoglazova[1,4], Alexander Duplinskiy[1,2,3], Vladimir Zavodilenko[1,2], Anton Losev[1,2], and Yury Kurochkin[1,2,3]**

*[1] QRate, 100 Novaya str., Skolkovo, Russian Federation*
*[2]NTI Center for Quantum Communications, National University of Science and Technology MISiS, 4 Leninsky prospekt, Moscow, Russian Federation*
*[3]Russian Quantum Center, 45 Skolkovskoye shosse, Moscow, Russian Federation*
*[4]Skolkovo Institute of Science and Technology, Bolshoy Boulevard 30, bld. 1, Moscow, Russian Federation*
**r.shakhovoy@goqrate.com*

**Abstract:** Direct phase modulation via optical injection is a newly developed method for coding the phase of a gain-switched laser, which meets high requirements placed on transmitters for quantum key distribution: compactness, low losses, compatibility with CMOS technologies, and the absence of undesirable effects leading to the side-channel information leakage. Despite the successful implementation and good prospects for the further development of this system, there is still a lack of theoretical investigations of this scheme in the literature. Here, for the first time, we perform its theoretical analysis. We study the influence of the spontaneous emission noise, examine the role of the gain non-linearity, and consider the temperature drift effect. The results obtained reveal that these phenomena significantly affect system performance. We have tried to formulate here practical instructions, which will help to take these features into account when elaborating and employing the optical-injection-based phase modulator.

## 1. Introduction

Nowadays, modulation of optical radiation is the technological basis of data transmission in telecommunications. In most cases, the amplitude and phase modulation of the laser radiation is performed via $LiNbO_3$-based modulators, the workhorse of optoelectronics for decades [1]. Despite the popularity and high efficiency of modulators of this type, they have some shortcomings, which are becoming more pronounced in the course of the incessant development of lightwave communication systems.

Although high-bandwidth $LiNbO_3$-based modulators have been mass-produced for a long time, they still have a high cost: the price of some models may be several times higher than the cost of standard telecommunication lasers and fiber-optic elements. Therefore, research and development of an effective and, at the same time, cheaper method of modulation are highly expedient from an economical point of view. Another important factors are 1) the quite large size of $LiNbO_3$-based phase modulators, whose characteristic lengths generally reach several centimeters, and 2) the high value of the half-wave voltage. These features limit their use in photonic integrated circuits (PICs) compatible with complementary metal-oxide-semiconductor (CMOS) technology [2]. It is worth noting that highly efficient hybrid Si-$LiNbO_3$ integrated Mach-Zehnder amplitude modulators with low half-wave voltage, small length (less than a centimeter), low loss, and high bandwidth have been recently developed [3, 4]. However, similar *phase* modulators have not yet been demonstrated.

Phase modulation is one of the most demanded ways to encode information in lightwave communications. Phase coding gains particular importance today due to the development of quantum key distribution (QKD) systems, where cryptographic keys are commonly encoded in the phase of attenuated laser pulses [5]. Following the worldwide trend towards miniaturization and development of PICs, developers of QKD systems are looking for efficient QKD transmitters on a chip. Generally, high demands are placed on such transmitters, including not only compactness, low losses, and compatibility with CMOS technologies, but also the absence of undesirable effects leading to the side-channel information leakage. Recently, Yuan et al. [6] demonstrated a direct phase modulation scheme, which seems to meet these requirements and is thus one of the possible candidates for replacing (at least in some applications) $LiNbO_3$-based phase modulators.

The proposed scheme of the direct phase modulation is based on the well-known phenomenon of phase locking in an optically-injected laser [7, 8]. It uses two semiconductor lasers, one of which (the pulse preparation laser) operates under the gain switching, whereas the second one (the phase preparation laser) operates in a quasi-steady-state regime and employs small perturbations in the pump current to control the phase between adjacent pulses of the slave laser. Actually, optical injection is a well-studied technique widely used to improve characteristics of directly modulated lasers [9], and optical phase locking has been known for a long time. Therefore, it is remarkable that such a method of direct phase modulation was invented only a few years ago. Recently, it has already been successfully implemented in QKD systems [10, 11].

Despite the successful implementation and good prospects for the further development of this system, there is still a lack of theoretical investigations of this scheme in the literature. In fact, Yuan et al. [6] provide only an intuitive picture to understand how an optical phase can be set to a secondary laser. However, for further development and improvement of this method, experimental data should be supported by theoretical analysis, which would relate various parameters of the system to its efficiency. This work is aimed to fill this gap. Here, we study the influence of the spontaneous emission noise, examine the role of the gain non-linearity, and consider the effect of the temperature drift. We believe that these three aspects of laser operation are the most important in the context of direct phase modulation and should be thus carefully treated.

In the next section, we provide a general description of a system for the direct phase modulation via optical injection. Then, the rate equation analysis will be used to consider the influence of master and slave laser properties on system performance.

## 2. General description of a system

The possible realization of a master-slave laser system for the direct phase modulation is schematically presented in Fig. 1. In this realization (so-called reflection style [9]), a pair of semiconductor lasers is connected via an optical circulator. (In a fiber-optic implementation, a polarization controller between the lasers is needed or, otherwise, laser pigtails and circulator must be made of a polarization-maintaining fiber.) The slave laser, which serves as a pulse generation source, operates in a gain-switched regime. The master laser, which acts as a phase preparation source, is directly modulated to produce quasi-steady-state emission with moderate perturbations as shown in Fig. 1, where the driving current of the phase preparation laser is modulated at times synchronized between the two consecutive pulses of the slave laser. By quasi-steady-state emission, we mean here that perturbations used to control the phase of the pulse preparation laser are small enough such that transients can be neglected.

In the absence of radiation from the master laser, the phases acquired by pulses of the gain-switched laser are random, inasmuch as phase correlations of the electromagnetic field are completely destroyed during the time between laser pulses due to the contribution of spontaneous recombination. When the master radiation is on, the slave cavity is not depleted between the pulses since it now contains photons from the phase preparation source, which will trigger the lasing of the next pulse. The phase of this pulse will not be random now but will be determined by an optical frequency of the master laser radiation. In fact, at sufficient optical power from the master laser, one may assume that the only field in the slave laser cavity is the master field, such that the phase between the slave laser pulses is determined predominantly by the phase evolution of this field. It should also be noted that there is no optical injection locking (in its usual sense) during the time interval between the slave laser pulses since the slave is not lasing there. So, the restriction imposed on the locked phase difference $\varphi_L$ between the master and the slave lasers [12] – $\varphi_L \in [-\pi/2 - \arctan\alpha, \pi/2 - \arctan\alpha]$ ($\alpha$ is the Henry factor [13]) – is irrelevant here.

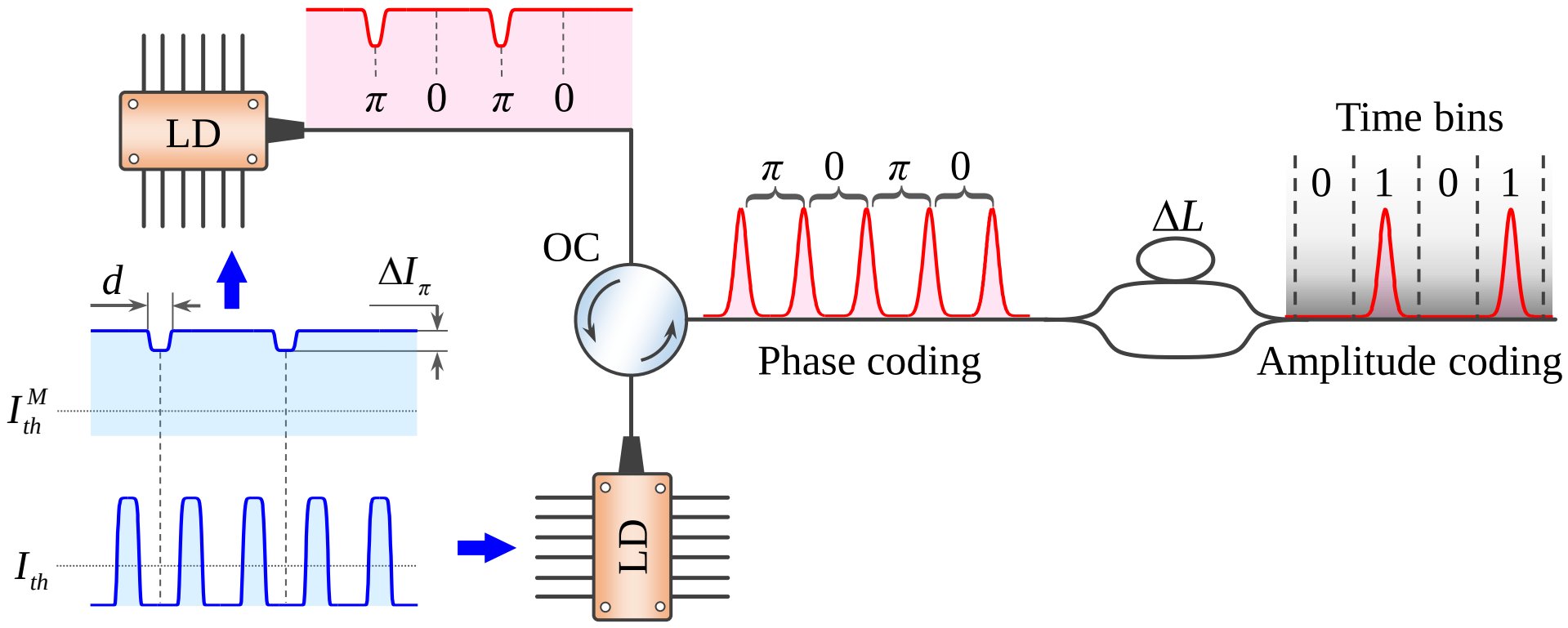

Fig. 1. Possible realization of a master-slave laser system for the direct phase modulation. $d$ is the duration of the master current perturbation; $\Delta I_\pi$ is the master current change needed for the phase modulation depth of $\pi$, $I_{th}^M$ and $I_{th}$ are threshold currents for the master and slave laser, respectively; $\Delta L$ is the single period delay line; LD stands for the laser diode, OC – the optical circulator.

Generally, the change of the pump current in a semiconductor laser causes the change of its carrier density, which, in turn, leads to the change of an optical frequency [13, 14]. Due to this, the change of the master driving current applied between the slave pulses allows transferring frequency change of the master radiation onto a phase difference between neighboring pulses of a gain switched laser. In Fig. 1, we assume for simplicity that modulation of the master is set to introduce the phase shift equal to 0 or $\pi$. The phase-encoded pulse train may be decoded, e.g., by an asymmetric Mach-Zehnder interferometer, whose delay line $\Delta L$ is chosen so that the corresponding delay time equals to the pulse repetition period of a slave laser. With such a system, the phase modulation is transferred onto amplitude modulation suitable for detection by conventional (or maybe single-photon) detectors. Details of possible experimental realizations of such a system (particularly, in application to QKD) can be found in the literature [6, 10, 11]; below, we focus on its theoretical analysis.

## 3. Rate equation analysis

### *3.1 The role of the gain compression factor*

The dynamics of the master-slave laser system can be effectively described by the set of coupled differential equations [15-18] for the master

$$\frac{dN^M}{dt}=\frac{I^M}{e}-\frac{N^M}{\tau_e^M}-\frac{Q^M}{\Gamma^M\tau_{ph}^M}G^M+F_N^M\,,$$

$$\frac{dQ^M}{dt}=\left(G^M-1\right)\frac{Q^M}{\tau_{ph}^M}+C_{sp}^M\frac{N^M}{\tau_e^M}+F_Q^M\,, \qquad (1)$$

$$\frac{d\varphi^M}{dt}=\frac{\alpha^M}{2\tau_{ph}^M}\left(G_L^M-1\right)+F_\varphi^M\,,$$

and for the slave laser

$$\frac{dN}{dt}=\frac{I}{e}-\frac{N}{\tau_e}-\frac{Q}{\Gamma\tau_{ph}}G+F_N\,,$$

$$\frac{dQ}{dt}=\left(G-1\right)\frac{Q}{\tau_{ph}}+C_{sp}\frac{N}{\tau_e}$$

$$+2\kappa_{ex}\sqrt{Q^M Q}\cos(\varphi-\varphi^M-\Delta\omega t)+F_Q\,, \qquad (2)$$

$$\frac{d\varphi}{dt}=\frac{\alpha}{2\tau_{ph}}\left(G_L-1\right)$$

$$-\kappa_{ex}\sqrt{\frac{Q^M}{Q}}\sin(\varphi-\varphi^M-\Delta\omega t)+F_\varphi\,,$$

where the superscript $M$ means “master” and distinguishes the variables and parameters of the master laser from the corresponding variables and parameters of the slave laser. Here $Q$ is the absolute square of the normalized electric field amplitude corresponding to the photon number inside the laser cavity and related to the output power by $P=Q(\varepsilon\hbar\omega_0/2\Gamma\tau_{ph})$, where $\hbar\omega_0$ is the photon energy ($\omega_0$ is the central frequency), $\varepsilon$ is the differential quantum output, $\Gamma$ is the confinement factor, $\tau_{ph}$ is the photon lifetime inside the cavity, and the factor $1/2$ takes into account the fact that the output power is generally measured from the only one facet. Onwards, $\varphi$ is the phase of the field, $N$ is the carrier number, $I$ is the pump current, $e$ is the absolute value of the electron charge, $\tau_e$ is the effective lifetime of the electron, the factor $C_{sp}$ corresponds to the fraction of spontaneously emitted photons that end up in the active mode, $\alpha$ is the linewidth enhancement factor (the Henry factor [13]), and the dimensionless linear gain $G_L$ is defined by $G_L=(N-N_{tr})/(N_{th}-N_{tr})$, where $N_{tr}$ and $N_{th}$ are the carrier numbers at transparency and threshold, respectively. The gain saturation [19] is included in Eqs. (1)-(2) by using the relation $G=G_L(1-\chi P)$ or, equivalently, by $G=G_L(1-\chi_Q Q)$, where $\chi$ is the gain compression factor [19-21] and $\chi_Q=\chi(\varepsilon\hbar\omega_0/2\Gamma\tau_{ph})$ is its dimensionless counterpart. (Note that equations for $Q$ and $N$ contain $G$, whereas equation for $\varphi$ contains the linear gain $G_L$.) The detuning between the master and the solitary slave lasers is denoted by $\Delta\omega$; the coupling between the lasers is determined by the parameter $\kappa_{ex}$, whose magnitude may be estimated from the relation $\kappa_{ex}\sim t_{MS}/\tau_L$, where $t_{MS}$ is the amplitude transmittance of the slave laser facet, from which the light of the master gets into the slave cavity (power transmittance is $T_{MS}=t_{MS}^2$), and $\tau_L$ is the time, during which the wave packet travels twice from facet to facet of the slave laser cavity (the so-called round-trip time). Finally, $F_{N,Q,\varphi}$ are Langevin forces defining fluctuations of the carrier number, the photon number, and the phase, respectively.

As shown in the previous section (see Fig. 1), phase modulation is governed by the change of the master laser pump current in the interval between the slave laser pulses. If the injected power is much higher than the spontaneous noise of the slave laser, the contribution of the pulse generation source to radiation between pulses can be neglected, and we can consider only the system (1). We assume further that the master laser produces quasi-steady-state emission during the phase encoding time, so we can write for the average (i.e., without noise) photon, carrier number, and the field phase the following steady-state equations:

$$\frac{I_s^M}{e}-\frac{N_s^M}{\tau_e^M}-\frac{Q_s^M}{\Gamma^M\tau_{ph}^M}\frac{N_s^M-N_{tr}^M}{N_{th}^M-N_{tr}^M}(1-\chi_Q^M Q_s^M)=0,$$
$$\left(\frac{N_s^M-N_{tr}^M}{N_{th}^M-N_{tr}^M}(1-\chi_Q^M Q_s^M)-1\right)\frac{Q_s^M}{\tau_{ph}^M}+C_{sp}^M\frac{N_s^M}{\tau_e^M}=0, \tag{3}$$
$$\frac{d\varphi^M}{dt}=\omega_s^M-\omega_0^M=\frac{\alpha^M}{2\tau_{ph}^M}\left(\frac{N_s^M-N_{tr}^M}{N_{th}^M-N_{tr}^M}-1\right),$$

where the subscript $s$ denotes the steady-state value. Unfortunately, the strict analytical solution of Eq. (3) is too cumbersome; however, we can use the fact that $C_{sp},\chi_Q<<1$ and expand the solution into the series of $C_{sp}$ and $\chi_Q$ leaving only the terms of the first order and neglecting other terms including cross terms proportional to $C_{sp}\chi_Q$. We will find then an approximate analytical solution:

$$Q_s\approx\frac{\Gamma\tau_{ph}}{e}(I_s-I_{th})\left(1+\frac{I_{th}(I_s-I_{tr})}{(I_s-I_{th})^2\Gamma}C_{sp}-\frac{\Gamma\tau_{ph}}{e}(I_{th}-I_{tr})\chi_Q\right),$$
$$N_s\approx\frac{I_{th}\tau_e}{e}\left(1-\frac{I_{th}-I_{tr}}{(I_s-I_{th})\Gamma}C_{sp}+\frac{\Gamma\tau_{ph}}{eI_{th}}(I_{th}-I_{tr})(I_s-I_{th})\chi_Q\right), \tag{4}$$
$$\omega_s-\omega_0\approx-\frac{\alpha}{2\Gamma\tau_{ph}}\frac{I_{th}}{I_s-I_{th}}C_{sp}+\frac{\alpha}{2e}\Gamma(I_s-I_{th})\chi_Q,$$

where $\omega_s$ is the stationary value of the optical frequency. We also used the relations $I_{th}=N_{th}e/\tau_e$ , $I_{tr}=N_{tr}e/\tau_e$ and omitted for brevity the superscript $M$ . As shown in Fig. 1, phase modulation is implemented due to different changes (perturbations) of the master driving current between different pairs of the slave pulses. So, the phase difference between different *pairs* of pulses can be written as $\Delta\Phi=d(\omega_{s1}-\omega_{s2})$ , where $d$ is the duration of the master current perturbation (see Fig. 1). Using Eq. (4) we find:

$$\Delta\Phi=\frac{\alpha d}{2\tau_{ph}}\frac{I_{th}(I_{s2}-I_{s1})}{(I_{s1}-I_{th})(I_{s2}-I_{th})}\frac{C_{sp}}{\Gamma}+\frac{\alpha d}{4\tau_{ph}}(I_{s2}-I_{s1})\frac{\varepsilon\hbar\omega_0}{e}\chi, \tag{5}$$

where we went back from $\chi_Q$ to $\chi$ . Denoting further $I_{s2}-I_{s1}=\Delta I$ , $I_{s1}\equiv I_s$ and $I_{s2}\equiv I_s+\Delta I$ , we can find from Eq. (5) the master current change $\Delta I_\pi$ needed for the phase modulation depth of $\Delta\Phi=\pi$ (see Fig. 1). Unfortunately, the corresponding analytical expression is quite bulky; therefore, we used the same approach as for the derivation of Eq. (4). We obtained:

$$\Delta I_\pi = \frac{2}{\chi}\frac{e}{\varepsilon\hbar\omega_0}\left(\frac{2\pi\tau_{ph}}{\alpha d} - \frac{C_{sp}}{\Gamma}\frac{I_{th}}{I_s - I_{th}}\right) + \frac{C_{sp}}{\Gamma}\frac{\alpha d}{2\pi\tau_{ph}} I_{th}. \quad (6)$$

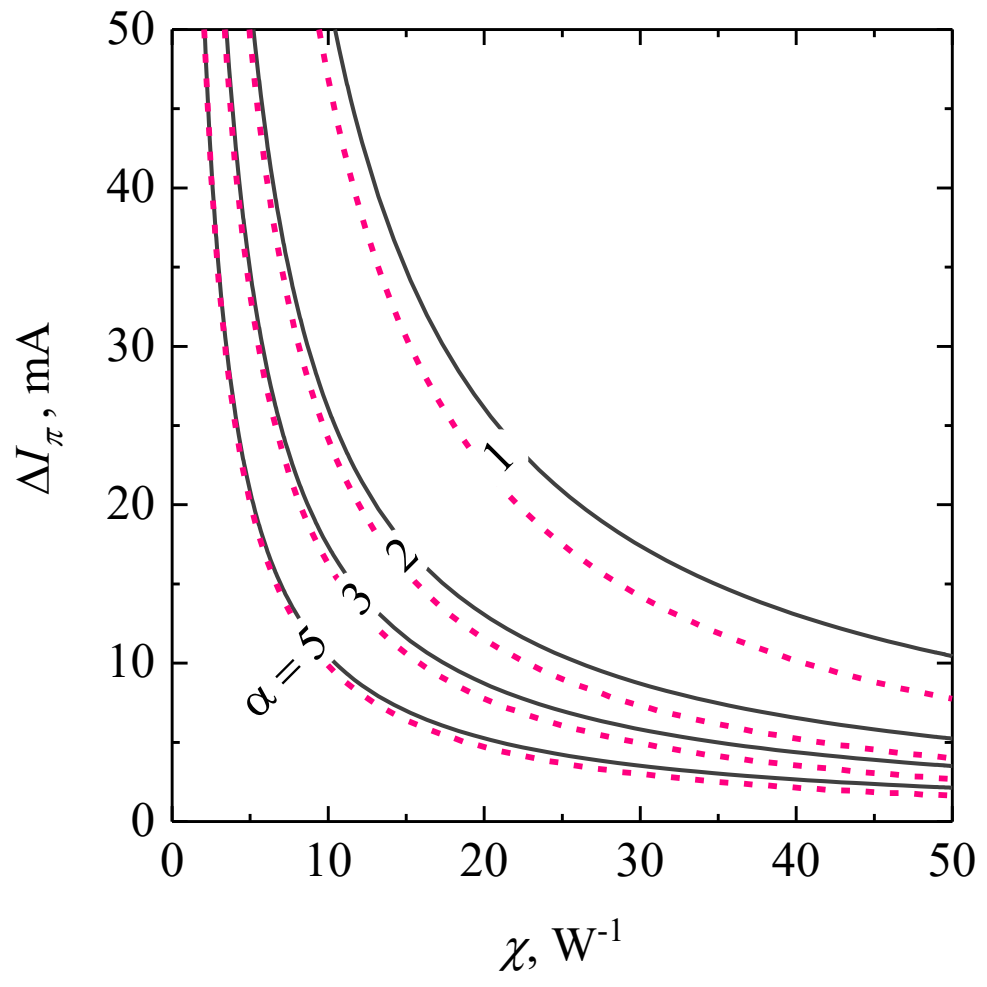

Fig. 2. The master current change $\Delta I_\pi$ needed for the phase modulation depth of $\Delta\Phi = \pi$ as a function of the gain compression factor $\chi$ at different values of the Henry factor $\alpha$. The dashed lines were found numerically from the solution of Eq. (3), whereas the solid lines correspond to Eq. (8).

Note that one can neglect with good accuracy the dependence of $\Delta I_\pi$ on $I_s$ when $I_s > 1.1 I_{th}$. This becomes clear if we take into account the fact that $\alpha d C_{sp} / 2\pi\Gamma\tau_{ph} \sim 10^{-2}$ for typical laser parameters (and for $d \sim 0.1$ ns), such that one can discard the second term in brackets of Eq. (6). Then, using the following estimation valid for typical semiconductor lasers:

$$\frac{C_{sp}}{8\pi^2\Gamma}\left(\frac{\alpha d}{\tau_{ph}}\right)^2 \frac{I_{th}}{e}\varepsilon\hbar\omega_0 \sim 10^{-1} d^2 \text{ W} \quad (7)$$

($d$ is in ns), we can also neglect the third term in Eq. (6). Such an approximation, in essence, corresponds to the absence of spontaneous radiation and is valid with very high accuracy in a broad range of $\chi$ and $d$ values in the case when the master laser operates well above threshold. Thus, we can estimate the master current change $\Delta I_\pi$ by the following simple relation:

$$\Delta I_\pi = \frac{4\pi}{\chi^M}\frac{e}{\varepsilon^M \hbar\omega_0^M}\frac{\tau_{ph}^M}{\alpha^M d}. \quad (8)$$

The dependence of $\Delta I_\pi$ on $\chi$, according to Eq. (8) in the range $\chi = 0..50$ W$^{-1}$ for different values of the Henry factor $\alpha$ is shown in Fig. 2 by the solid line. Corresponding laser parameters are listed in Table 1; the duration of the master current perturbation was put to $d = 0.1$ ns. Dashed lines on the figure correspond to the same dependences found numerically from the system (3) with the same set of laser parameters (the pump current was put to $I_s = 30$ mA). When decreasing $\alpha$, the difference between exact and approximate dependences becomes more pronounced: Eq. (8) provides an overestimated value for $\Delta I_\pi$. However, at $\alpha > 3$ both dependencies are in good agreement, particularly at smaller values of $\chi$. So, for typical values of the linewidth enhancement factor ($\alpha = 3..6$), Eq. (8) can be used without significant restrictions. Note here that the selected range of changes for $\chi$ corresponds to the typical range for semiconductor lasers and can be estimated from the following formula: $\chi = 2\chi_q \Gamma \tau_{ph} / (\varepsilon \hbar \omega_0 V)$, where $V$ is the volume of the active region (which is generally in the range $10^{-16} - 10^{-15}$ $m^3$) and $\chi_q = \chi_Q V$ is the photon-density-related gain compression factor (which is generally in the range $10^{-23} - 10^{-22}$ $m^3$ [19, 22]) defined by $G = G_L(1 - \chi_q q)$, where $q = Q/V$ is the photon density. One can see that $\chi$ depends on $V$ and may reach a few dozens of $W^{-1}$ for lasers with the small active region.

**Table 1. Laser Parameters Used in Simulations**[a]

| Parameter | Value |
|---|---|
| Photon lifetime $\tau_{ph}$, ps | 1.0 |
| Electron lifetime $\tau_e$, ns | 1.0 |
| Quantum differential output $\varepsilon$ | 0.3 |
| Transparency carrier number $N_{tr}$ | $4.0 \times 10^7$ |
| Threshold carrier number $N_{th}$ | $5.5 \times 10^7$ |
| Spontaneous emission coupling factor $C_{sp}$ | $10^{-5}$ |
| Confinement factor $\Gamma$ | 0.12 |
| Linewidth enhancement factor $\alpha$ | 5 |
| Gain compression factor $\chi$, $W^{-1}$ | 30 |
| Master-slave detuning $\Delta\omega/2\pi$, Hz | 0 |

[a]The listed parameters were used both for the slave and master lasers

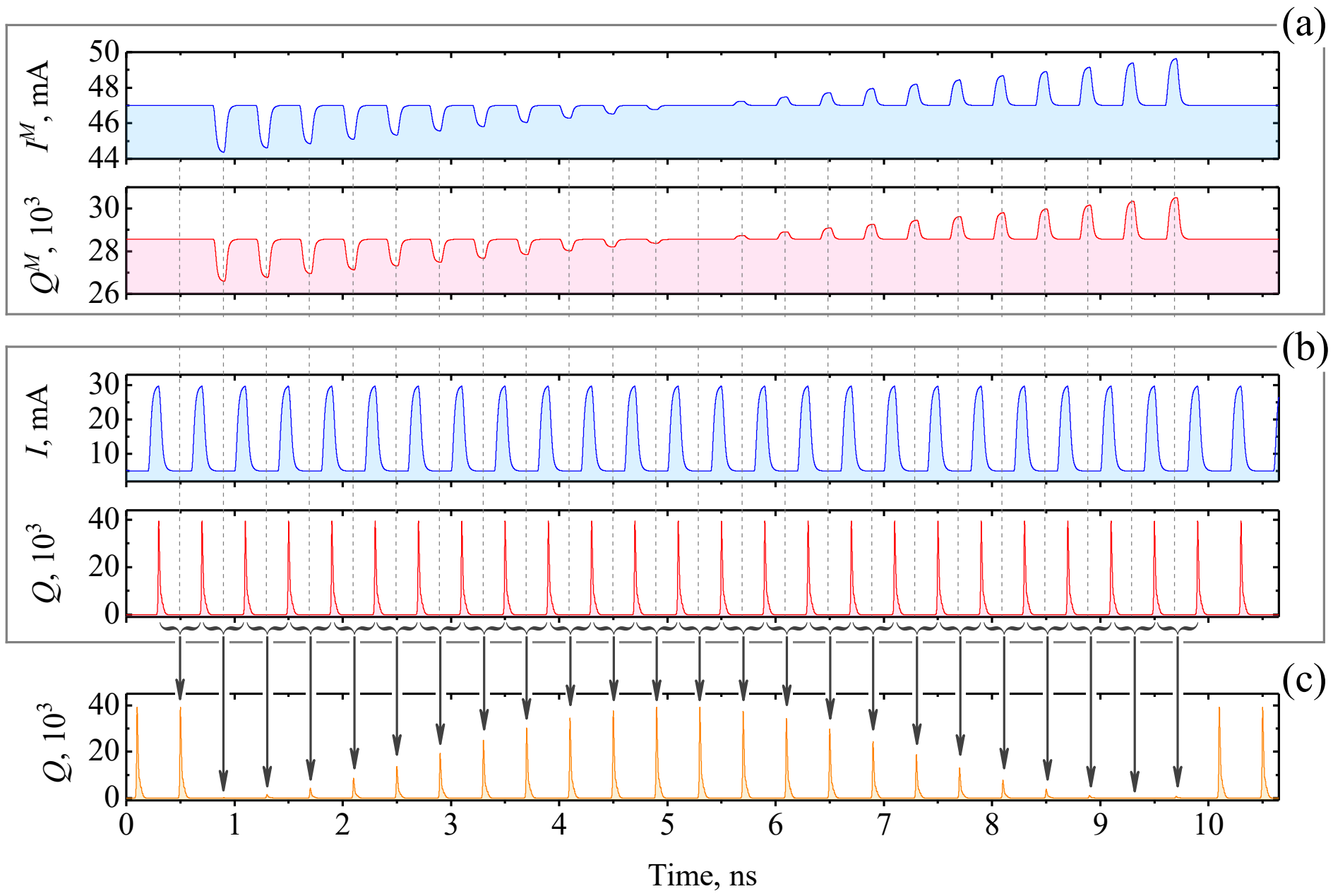

Fig. 3. Simulations of the phase modulation via optical injection without spontaneous noise: (a) the modulation current (blue line) and the output intensity (red line) of the master laser; (b) the modulation current (blue pulses) and the output intensity (red pulses) of the slave laser; (c) the result of the interference of the slave pulse train with itself shifted by one modulation period.

To demonstrate the phase modulation efficiency, we solved the system (1)-(2) without spontaneous noise with the master modulation current pattern shown in panel (a) of Fig. 3 by the blue line; the corresponding output intensity of the master laser is shown just below the modulation signal by the red line. The modulation current and the output intensity of the slave laser are shown in panel (b). In panel (c), we put the result of the interference of the slave pulse train with itself shifted by one modulation period. For simulations, we used laser parameters from Table 1; the coupling coefficient $\kappa_{ex}$ was set to 10 GHz. Here, according to numerical simulations, $\Delta I_{\pi} \approx 2.7$ mA, which is in good agreement with Eq. (8). Indeed, one can see that linear variation of the master current from $-\Delta I_{\pi}$ to $\Delta I_{\pi}$ provides an interference fringe from $-\pi$ to $\pi$.

### *3.2 The influence of temperature*

As we noted earlier, the master laser in the modulation scheme under consideration is assumed to operate in a quasi-steady-state regime; therefore, we neglected above all issues related to transients. Such a simplification, however, is not always applicable. Unwanted effects could, in principle, be observed when encoding pulse trains, whose length is comparable to (or less than) the so-called thermal rise-time (see Appendix) of the master laser. The use of pulse trains consisting of tens to thousands of laser pulses is quite natural for QKD; moreover, some protocols, e.g., round-robin-differential-phase-shift (RRDPS) QKD protocol [23] inherently employs short pulse trains. So, it is highly reasonable to follow the influence of long-term temperature-related transients on the phase modulation of such pulse trains.

Generally, when an electric current passes through a laser diode, an increase in temperature within the laser volume is observed. This phenomenon is related to non-radiative recombination processes as well as to the reabsorption of the generated radiation. Another possible effect is the Joule heat generated at the electric contacts and in the heat sink. What is important for us is that the temperature rise within the laser active layer leads to the shift of the lasing frequency, which is caused by changes in the size of the Fabry-Perot cavity (or in the size of the Bragg grating pitch in the case of DFB lasers), and by changes in the value of the refractive index of the cavity material.

Several models have been proposed to estimate the transient temperature rise in the active layer [24-26]. A simple model has been developed by Engeler and Garfinke [24], who assumed a steady heat injection from the junction region to a thick substrate layer connected to a heat sink equipped with a thermoelectric cooler. Although this model is difficult to apply, it can be reformulated in terms of effective phenomenological parameters, which may be estimated or even measured experimentally. In the Appendix, we provide a detailed discussion of this approach, which leads to the following equation:

$$\frac{d\Delta T}{dt} = -\frac{\Delta T}{\tau_h} + \frac{1.24}{\lambda}\frac{r_h}{\tau_h}(1-\varepsilon)(I - I_b), \tag{9}$$

where $\Delta T = T - T_0$ is the deviation of the active layer temperature from the steady-state value $T_0$, which is assumed to achieve at $I = I_b$, where $I_b$ is the bias current. In Eq. (9), $r_h$ is the thermal resistance between the laser active layer and the heat sink, $\tau_h$ is the thermal rise-time, $\lambda$ is the lasing wavelength, and, as above, $\varepsilon$ is the differential quantum output.

The thermal resistance can be estimated from the following relation: $r_h = l/(2kA)$, where $l$ is the width of the semiconductor material between the active layer and the heat sink, $k$ is its thermal conductivity, and $A$ is the area of the interface between the active layer and the clamping layers. Assuming that the thickness $l$ of the substrate is 1-2 μm, and putting $A \sim 10^3$ μm$^2$ (the length and the width of the active layer are 500 and 2 μm, respectively), we obtain for both InP and GaAs substrates ( $k = 68$ and $55\ \mathrm{W\,m^{-1}K^{-1}}$ 55 W m$^{-1}$K$^{-1}$, respectively [27]): $r_h \sim 10$ K W$^{-1}$. The value of $r_h$ can be measured experimentally with the thermal resistance analysis by induced transient method (the so-called TRAIT method) [28-30]. This method allows experimental evaluation of the total thermal resistance $r_h$ of a semiconductor device and its assembling structure with a spatial resolution. For conventional InGaAsP-based telecom lasers, the total thermal resistance $r_h$ was measured to be of the order of 10-100 K/W depending on the packaging, in agreement with our estimation.

For the measurement of the thermal rise-time, Abdelkader et al. [31] studied the relation of the threshold current with temperature under pulsed operation and proposed a method of measuring $\tau_h$. This technique has been successfully applied to several telecom laser diodes [32] and $\tau_h$ was measured to be of the order of 10-40 ns.

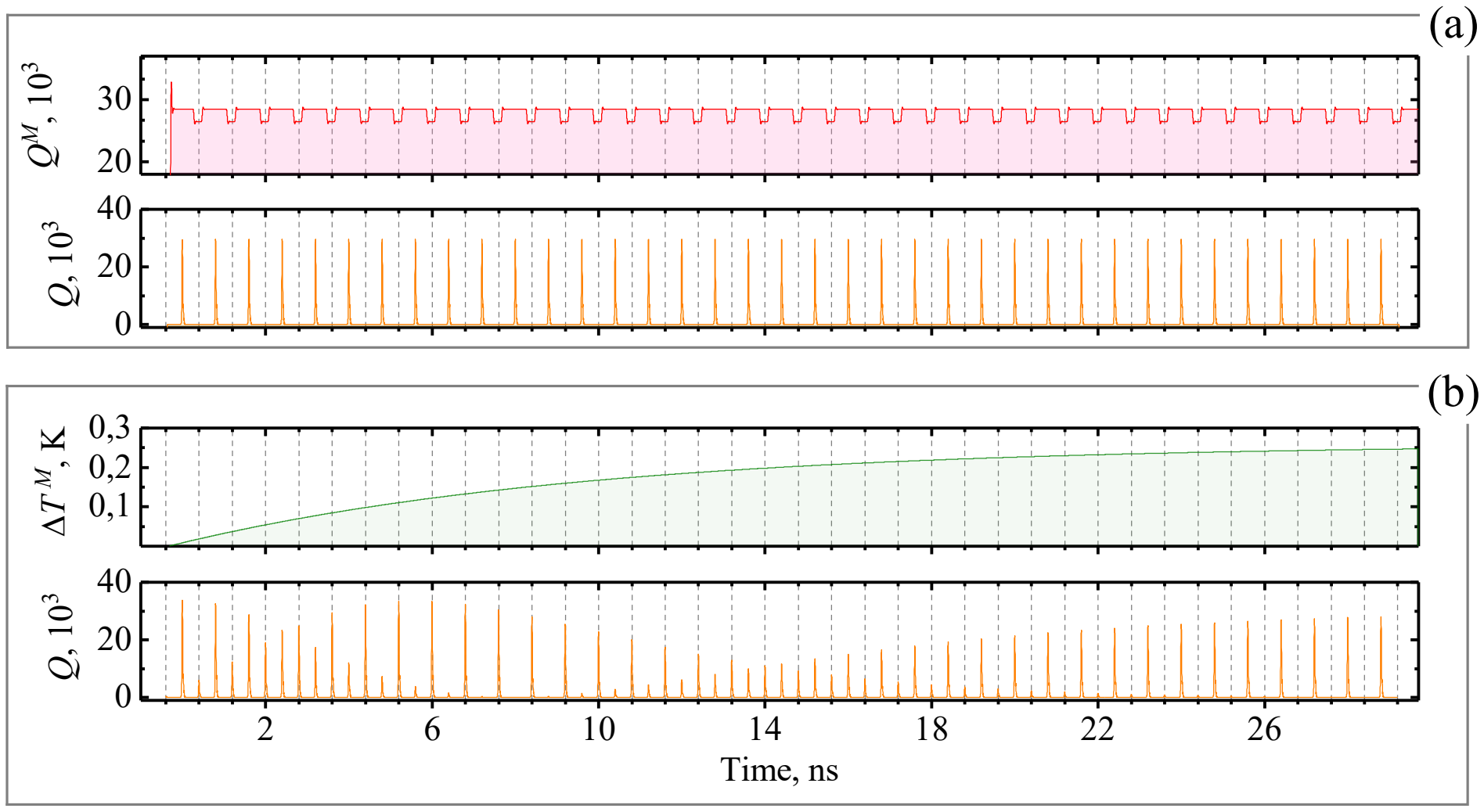

Fig. 4. Simulations of the phase modulation via optical injection without spontaneous noise: (a) the output intensity of the master laser ( $Q^M$ ) and the corresponding result of the interference ( $Q$ ) of the slave pulse train with itself shifted by one modulation period without taking into account the temperature effect; (b) the temperature drift of the master laser after switching on (green line) and the corresponding result of the interference of the slave pulse train with itself as in (a).

Finally, one can show that the temperature change does not affect the amplitude of the laser field (at least, at the first-order approximation), such that only the phase rate equation will be modified. The latter will have the form (without spontaneous noise term):

$$\frac{d\varphi}{dt} = \frac{\alpha}{2\tau_{ph}}\left(G_L - 1\right) - \mu_\omega \Delta T, \tag{10}$$

where $\mu_\omega$ is the temperature coefficient of frequency, which is related to the temperature coefficient of the wavelength $\mu_\lambda$ as follows: $\mu_\omega = \omega^2 \kappa_\lambda / 2\pi c$. It is measured experimentally that $\mu_\lambda \approx 0.06 - 0.07$ nm/K for GaAs-based lasers in 800 nm band, and $\mu_\lambda \approx 1$ nm/K for InP-based lasers in 1550 nm band [33]. So, the temperature coefficient of frequency for InP-based 1550 nm lasers is $\mu_\omega / 2\pi \sim 10$ GHz/K. The temperature change $\Delta T$ is related to the modulation current by Eq. (9).

Modifying the phase rate equation for the master laser according to Eq. (10) we can follow the influence of temperature on the direct phase modulation. Corresponding simulations are shown in Fig. 4. In panel (a), we presented the output intensity of the master laser ( $Q^M$ ) and the corresponding result of the interference of the slave pulse train with itself shifted by one modulation period. In panel (b), we showed the same interference, but with the temperature drift of the master laser to be turned on according to Eq. (9); the temperature drift itself, $\Delta T^M$, is shown above the resulting pulse train by the green line. Thermal properties of the master laser were specified by the following set of parameters: $\mu_\omega / 2\pi = 10$ GHz/K, $\tau_h = 10$ ns, and $r_h = 10$ K/W (the lasing wavelength appearing in Eq. (9) was set to 1550 nm). Other laser parameters were again taken from Table 1. The spontaneous noise was not included in this simulation.

One can see that at the onset of lasing, the master laser exhibits an apparent increase in temperature, which, according to the above consideration, leads to the time-dependent frequency shift, which, in turn, causes the drift of the phase change $\Delta\Phi$ between the pulses of the slave laser. It is clear from Fig. 4 that stable phase modulation occurs only after ~20 ns, when the master achieves a new stationary temperature, whereas at the beginning of the master lasing, the phase modulation cannot be easily controlled by identical perturbations and, obviously, the more comprehensive scheme is required.

### *3.3 The influence of spontaneous noise*

Finally, let us follow the influence of the spontaneous noise on the phase modulation efficiency. For this, one needs to solve the system (1)-(2) with stochastic terms. An approach to finding an explicit form of Langevin terms can be found elsewhere [34]; here, we just list them without derivation:

$$\begin{aligned} F_Q dt &= 2\sqrt{\frac{C_{sp}\bar{N}\bar{Q}}{2\tau_e}}\left(\cos\bar{\varphi}dW^A + \sin\bar{\varphi}dW^B\right), \\ F_\varphi dt &= \sqrt{\frac{C_{sp}\bar{N}}{2\bar{Q}\tau_e}}\left(\cos\bar{\varphi}dW^B - \sin\bar{\varphi}dW^A\right), \\ F_N dt &= -2\sqrt{\frac{C_{sp}\bar{N}\bar{Q}}{2\tau_e}}\left(\cos\bar{\varphi}dW^A + \sin\bar{\varphi}dW^B\right) \\ &\quad + \sqrt{\frac{2\bar{N}}{\tau_e}}dW^C, \end{aligned} \tag{11}$$

where $W^A$, $W^B$, and $W^C$ are independent Wiener processes. (The Langevin terms for the master laser are defined in the same way, but with corresponding laser parameters and other three independent Wiener processes.) The bar over $Q$, $\varphi$, and $N$ indicates that amplitudes of the noise terms are still time-dependent, albeit they can be considered constant during the fluctuation averaging time. Generally, $\bar{N}$, $\bar{Q}$, $\bar{\varphi}$ and $\bar{N}^M$, $\bar{Q}^M$, $\bar{\varphi}^M$ are just a solution of the system (1)-(2) without stochastic terms. With Eq. (11), we can perform numerical integration of the system (1)-(2), e.g., with the use of the Euler-Maryama method [35], which is the simplest time discrete approximation used for integration of stochastic differential equations.

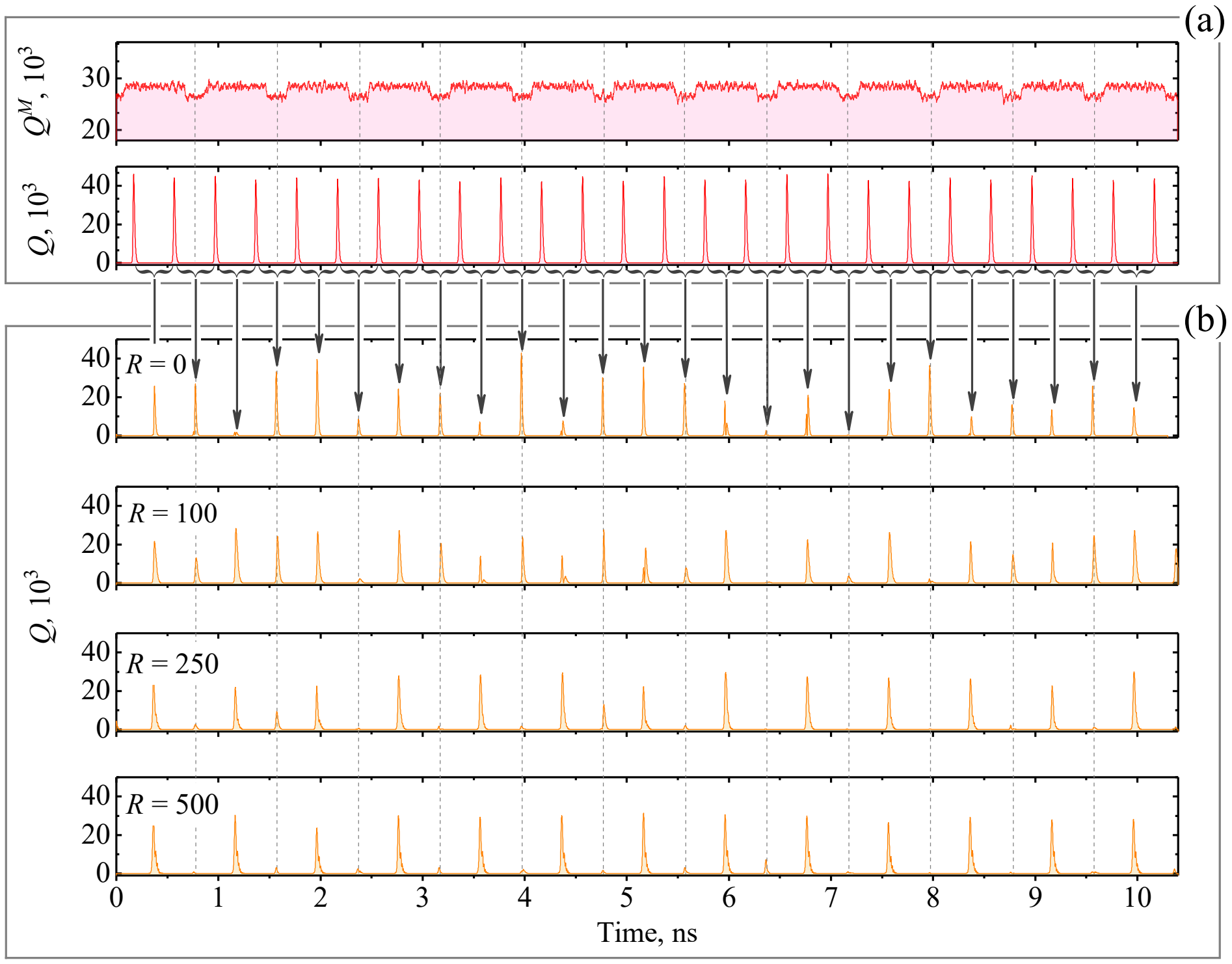

Fig. 5. Simulations of the direct phase modulation with spontaneous noise: (a) the output intensity of the master (above) and the slave (below) lasers; (b) the result of the interference of the slave pulse train with itself shifted by one modulation period at different values of a parameter $R$.

Simulations of the phase modulation via optical injection with the spontaneous noise are shown in Fig. 5. For simplicity, it is assumed in these simulations that temperature transients on the master laser are over. In panel (a), we showed the output intensity of the master and slave lasers. The master modulation was set so that adjacent pairs of the slave laser pulses interfered in the opposite way. In panel (b), we put the result of the interference at different values of the coupling coefficient $\kappa_{ex}$. However, it seems reasonable to use in this comparison the averaged ratio $R$ between the magnitude of the master contribution to the slave laser phase, $\kappa_{ex}\sqrt{Q^M/Q}$, and the diffusion coefficient for phase fluctuations given by the squared amplitude of the corresponding Langevin force, $C_{sp}\bar{N}/(2\bar{Q}\tau_e)$:

$$R = \kappa_{ex}\frac{2\tau_e}{C_{sp}}\frac{1}{Z}\sum_{j=1}^{Z}\frac{\sqrt{\bar{Q}_j^M\bar{Q}_j}}{\bar{N}_j}, \qquad (12)$$

where $\bar{N}_j \equiv \bar{N}(t_j)$, $\bar{Q}_j^M \equiv \bar{Q}^M(t_j)$, $\bar{Q}_j \equiv \bar{Q}(t_j)$, and $Z = T/\Delta t$, where $T$ is the pulse repetition period on the slave laser and $\Delta t$ is the integration step. The quantity $R$ defined in such a way serves as an evaluation parameter providing information on the value of $\kappa_{ex}$ (or the output master intensity $Q^M$, which is the same in this context) needed for efficient, i.e., resistant to internal noises, phase modulation. One can see that at $R = 0$ (without optical injection) interference of neighboring laser pulses is random due to their uncorrelated phases.

At small values of $R$ ($R < 500$), the spontaneous noise spoils the result of the interference, such that the phase modulation is hardly possible in this regime. We revealed that efficient phase modulation occurs at values of $R$ exceeding 500; this is clearly seen from the last simulation in panel (b) of Fig. 5.

## 4. Discussion

The first important result of the above theoretical analysis is the strong dependence of the phase modulation efficiency on the gain saturation. Indeed, Eq. (8) demonstrates that the gain compression factor $\chi$ plays a key role in phase modulation via optical injection. Already for the $\alpha = 5$ curve in Fig. 2, the master current change $\Delta I_\pi$ exceeds 10 mA at $\chi < 5$ W$^{-1}$, which can be accompanied by certain problems. In fact, such a value of driving current variations can hardly be regarded as a small perturbation and can already lead to the occurrence of transients. If $\chi$ is very small, the value of $\Delta I_\pi$ tends to become very high, such that the phase modulation becomes unpractical. In other words, to obtain a robust device, one needs to use the master laser with sufficiently high gain non-linearity.

One can also see from Eq. (8) that in addition to a high gain compression factor, it is preferable to choose a laser with a low quality factor (a small photon lifetime), a high Henry factor (high chirp), and a shorter wavelength in order to achieve a smaller value of $\Delta I_\pi$. Particularly, the DFB laser with a broader spectrum and a higher value of a threshold current (among similar lasers in a batch) should exhibit better performance. The latter becomes clear from the following relation: $I_{th} = I_{tr} + e\left(\tau_{ph}\tau_e v_g \Gamma\, \partial g/\partial N\right)^{-1}$, demonstrating that the threshold current is inversely proportional to the photon lifetime. (Here, $v_g$ is the group velocity of light and $\Gamma g = g_{opt}$ is the optical gain of the semiconductor medium.)

We also revealed that small temperature variations may significantly affect the phase modulation. It is important to note that standard thermal stabilization of a laser implemented, e.g., via a thermoelectric cooler, does not prevent this feature. The result shown in Fig. 4 demonstrates that after a long (tens of nanoseconds) “downtime” of the master laser, it could be problematic to achieve a stable phase modulation at the onset of master’s lasing since one will always deal with the temperature drift. This is also true even for the case when the master laser generates the trains of pulses (let say, M-shape pulses as, e.g., in [6]) instead of a long pulse shown in Fig. 4, and one will observe the same effect at the beginning of the pulse train. Therefore, in such a phase modulator, it is desirable to keep the master laser on (either under quasi-steady-state or under continuous pulsation) to avoid undesirable effects of the phase drift.

Another theoretical result shown in Fig. 5 indicates that the stable phase locking of the slave laser operating under the gain switching can be achieved only at a certain threshold output power from the master. Note that this is true even at zero detuning. This is an important difference from the optical injection scheme, in which both the slave and the master lasers operate in a CW mode. As mentioned above, this is due to the fact that at a low optical power entering the slave laser from the master, the influence of spontaneous noise on the phase dominates over the master's contribution. Therefore, for satisfactory operation of the phase modulator under consideration, it is preferable to use a master with relatively high output power. The required light intensity can be estimated from the value of the parameter $R$ defined by Eq. (12). It should be taken into account that the modulator may introduce coding errors at small $R$ values ($R < 500$), which will increase the bit error rate. Note, however, that significant optical power from the master laser may degrade the system performance. This can be the intense scattered light yielding the background power between the pulses as well as an undesirable effect of parasitic burning of charge carriers, which may lead to significant suppression of slave laser emission. (The latter effect is beyond the scope of the present study and will be discussed elsewhere.)

## 5. Conclusions

In this work, we performed a theoretical analysis of a master-slave laser system for direct phase modulation. The results obtained reveal that such a system has several features, which should be taken into account when elaborating and using the optical-injection-based phase modulator. We found that the gain non-linearity plays a crucial role in phase modulation, increasing its efficiency. To make our theoretical treatment convenient for practical use, we derived an approximate analytical formula (Eq. (8)) relating the value of the master current perturbation $\Delta I_{\pi}$ with experimentally measurable laser parameters. Also, we revealed that the laser system with the direct phase modulation is sensitive to temperature transients; therefore, it is important to choose an operation mode, which minimizes the temperature drift. Finally, we showed that the spontaneous noise may affect the efficiency of the direct phase modulation; therefore, to avoid the appearance of coding errors, the master output intensity should not fall below a certain critical value.

To the best of our knowledge, these features of the direct phase modulation have not been previously considered in the literature. We believe that obtained theoretical results will help to improve the efficiency of schemes for the direct phase modulation, which have already started to appear in laboratories, and probably shed light on the experimental effects the developers face with.

## Appendix

Following the model of Engeler and Garfinke [24], let us consider the highly simplified laser structure consisting of just three layers: the active layer and the two clumping $n$- and $p$-type layers of a much greater thickness. We will treat the active layer as the main heat source and will assume for simplicity that the heat flows equally into the clamping layers, such that the heat power flux into the $n$-type equals to the flux into the $p$-type material and is $P/2A$, where $P$ is the heat power dissipated in the active layer, and $A = wL$ is the area of the interface between the layers; here, $L$ and $w$ are the length and the width of the active layer (in a gain-guided laser $w$ approximately corresponds to the width of the stripe contact). The latter approximation implicitly assumes that the thermal conductivity of both $n$- and $p$-type materials is the same. To make the problem symmetric, we will also assume that both passive layers are connected to heatsinks, such that we may consider the problem of the heat flow into the $n$-material only. Finally, denote the width of the $n$-type passive layer as $l$ and assume that the $x$ axis is perpendicular to the layer surface putting $x = 0$ at the boundary between the heatsink and the $n$-layer. The problem may be thus considered one-dimensional, the heat transfer equation may be then written as follows:

$$\rho C \frac{\partial}{\partial t} T - \frac{\partial}{\partial x}\left( k \frac{\partial T}{\partial x} \right) = -\frac{P_J}{V}, \tag{13}$$

where $T$ is the temperature, $\rho$, $C$ and $k$ are the specific mass, the specific heat capacity, and the thermal conductivity of the semiconductor material, and, finally, $P_J$ is the Joule heat power generated in the laser volume $V$. The active layer heating is included by specifying the boundary condition at $x = l$:

$$k \left. \frac{\partial T}{\partial x} \right|_{x=l} = \frac{P}{2A}. \tag{14}$$

The other necessary boundary conditions are that $T = T_0$ at $x = 0$ and that $T = T_0$ at all $x$ at $t = 0$. Assuming further that the heat pulse is rectangular (such that $P$ does not depend on time) and its duration is $t$, we can solve Eq. (13) and obtain for the temperature difference $\Delta T = T - T_0$ at the end of the heat pulse:

$$\Delta T(t) = \frac{P}{kA\sqrt{\pi}}\sqrt{Dt}\left\{1 + 2\sqrt{\pi}\sum_{n=1}^{\infty}(-1)^n \operatorname{ierfc}\left(\frac{nl}{\sqrt{Dt}}\right)\right\}, \tag{15}$$

where we neglected the Joule heat on the right side of Eq. (13) and introduced the heat diffusion coefficient:

$$D = \frac{k}{\rho C}. \tag{16}$$

In Eq. (15), $\operatorname{ierfc}(x)$ is the first iterated integral of the complementary error function:

$$\operatorname{ierfc}(z) = \frac{1}{\sqrt{\pi}}e^{-z^2} - z\operatorname{erfc}(z). \tag{17}$$

Let us perform the following variable replacement $Dt/l^2 = p$ and introduce the function

$$\frac{kA\sqrt{\pi}}{Pl}\Delta T(p) = y(p) = \sqrt{p}\left\{1 + 2\sqrt{\pi}\sum_{n=1}^{\infty}(-1)^n \operatorname{ierfc}\left(\frac{n}{\sqrt{p}}\right)\right\}. \tag{18}$$

It is easy to show (e.g., by fitting) that the function $y(p)$ can be well approximated by the function

$$f(p) = 0.87\left(1 - e^{-3.29p}\right), \tag{19}$$

whence we have

$$\Delta T(t) \approx r_h P\left(1 - e^{-t/\tau_h}\right), \tag{20}$$

where

$$r_h = \frac{0.87\,l}{kA\sqrt{\pi}} \approx \frac{1}{2k}\frac{l}{A} \tag{21}$$

corresponds to the thermal resistance between the laser active layer and the heat sink, and

$$\tau_h = \frac{l^2}{3.9D} \tag{22}$$

is called the thermal rise-time.

The time dependence of the temperature difference $\Delta T(t)$ given by Eq. (20) can be considered as a solution to the following phenomenological equation:

$$\frac{d\Delta T}{dt} = -\frac{\Delta T}{\tau_h} + \frac{r_h P}{\tau_h}, \tag{23}$$

which defines the dynamics of the active layer temperature. The physical content of Eq. (23) is straightforward. The first term corresponds to the decrease of $\Delta T$ in the absence of the heat flux, and the rate of this decrease equals to $\tau_h^{-1}$. The second term governs the deviation of the active layer temperature from $T_0$ due to the heat power $P$ generated in it. Under steady-state, $d\Delta T/dt = 0$, the deviation from the heat sink temperature is given by $\Delta T = r_h P$.

Probably, the derivation of Eq. (23) presented above looks some unnatural; however, it provides a simple relation of phenomenological parameters $r_h$ and $\tau_h$ to material constants of the laser. Note also that numerical coefficients in Eqs. (21) and (22), which come from fitting parameters, have clear physical meaning. Thus, the factor $1/2$ in the definition of $r_h$ corresponds to the fact that, according to the above consideration, the power $P$ dissipated in the active layer flows equally into both clamping layers, such that the effective surface area is doubled and consequently, the resistance is halved. The factor 3.9 in Eq. (22) can be understood as follows. One can see that for $t = 3.9\tau_h$ the diffusion length given by $\sqrt{Dt}$ equals to the length of the sample. So, at the time $t = 3.9\tau_h$ the temperature difference is $\Delta T = 0.98 r_h P$, i.e., the steady-state temperature is almost reached, and any further increase in the length of the heat pulse causes no further temperature rise.

The heat is produced in the active layer due to the current flow or, more precisely, by the fraction of a current, which does not produce the output light. Introducing the voltage across the active layer as $V_g$, the heat power becomes

$$P = V_g I(1-\varepsilon), \tag{24}$$

where $\varepsilon$ is the differential quantum output and $I$ is the injection current. Assuming for simplicity that $V_g$ approximately equals to the band gap $E_g$ of the active layer material (in electronvolts) and using the relation [36] $E_g \approx 1.24/\lambda$, where $\lambda$ is the lasing wavelength in μm, we obtain for the 1.55 μm laser: $P = 0.8I(1-\eta)$.

The boundary conditions used above to solve Eq. (13) assume that the temperature of the active layer and of the heat sink is the same when there is no pumping. Obviously, the temperature of the active layer will be higher than that of the heat sink at some pre-pumping current $I_b$. However, let us put $\Delta T$ to zero if $I = I_b$ and treat $T_0$ as the steady-state temperature of the active layer. The change of the injection current will then cause the change of $\Delta T$, which can be positive or negative depending on whether the current value $I$ is greater or less than $I_b$. Equation (23) will be then written in the form of Eq. (9) in the main text.

## Funding

Russian Science Foundation (Grant No. 17-71-20146)

## Disclosures

The authors declare no conflicts of interest

## References

1. G. P. Agrawal, *Fiber-Optic Communications Systems* (John Wiley & Sons, New York, 2002), p. 546.
2. R. Baets and B. Kuyken, "High speed phase modulators for silicon photonic integrated circuits: a role for lithium niobate?," Advanced Photonics **1**, 1-3 (2019).
3. M. He, M. Xu, Y. Ren, J. Jian, Z. Ruan, Y. Xu, S. Gao, S. Sun, X. Wen, L. Zhou, L. Liu, C. Guo, H. Chen, S. Yu, L. Liu, and X. Cai, "High-performance hybrid silicon and lithium niobate Mach–Zehnder modulators for 100 Gbit s−1 and beyond," Nat. Photonics **13**, 359-364 (2019).
4. C. Wang, M. Zhang, X. Chen, M. Bertrand, A. Shams-Ansari, S. Chandrasekhar, P. Winzer, and M. Lončar, "Integrated lithium niobate electro-optic modulators operating at CMOS-compatible voltages," Nature **562**, 101-104 (2018).
5. N. Gisin, G. Ribordy, W. Tittel, and H. Zbinden, "Quantum cryptography," Rev. Mod. Phys. **74**, 145-195 (2002).
6. Z. L. Yuan, B. Fröhlich, M. Lucamarini, G. L. Roberts, J. F. Dynes, and A. J. Shields, "Directly Phase-Modulated Light Source," Phys. Rev. X **6**, 031044 (2016).

7. R. Lang, "Injection locking properties of a semiconductor laser," IEEE J. Quantum. Elect. **18**, 976-983 (1982).
8. S. Kobayashi and T. Kimura, "Optical phase modulation in an injection locked AlGaAs semiconductor laser," IEEE J. Quantum. Elect. **18**, 1662-1669 (1982).
9. E. K. Lau, L. J. Wong, and M. C. Wu, "Enhanced Modulation Characteristics of Optical Injection-Locked Lasers: A Tutorial," IEEE J. Sel. Top. Quant. **15**, 618-633 (2009).
10. G. L. Roberts, M. Lucamarini, J. F. Dynes, S. J. Savory, Z. L. Yuan, and A. J. Shields, "A direct GHz-clocked phase and intensity modulated transmitter applied to quantum key distribution," Quantum Science and Technology **3**, 045010 (2018).
11. T. K. Paraïso, I. De Marco, T. Roger, D. G. Marangon, J. F. Dynes, M. Lucamarini, Z. Yuan, and A. J. Shields, "A modulator-free quantum key distribution transmitter chip," npj Quantum Information **5**, 42 (2019).
12. F. Mogensen, H. Olesen, and G. Jacobsen, "Locking conditions and stability properties for a semiconductor laser with external light injection," IEEE J. Quantum. Elect. **21**, 784-793 (1985).
13. C. Henry, "Theory of the linewidth of semiconductor lasers," IEEE J. Quantum. Elect. **18**, 259-264 (1982).
14. B. R. Bennett, R. A. Soref, and J. A. D. Alamo, "Carrier-induced change in refractive index of InP, GaAs and InGaAsP," IEEE J. Quantum. Elect. **26**, 113-122 (1990).
15. G. H. M. v. Tartwijk and D. Lenstra, "Semiconductor lasers with optical injection and feedback," Quantum. Semicl. Opt. **7**, 87-143 (1995).
16. C. Henry, N. Olsson, and N. Dutta, "Locking range and stability of injection locked 1.54 μm InGaAsp semiconductor lasers," IEEE J. Quantum. Elect. **21**, 1152-1156 (1985).
17. I. Petitbon, P. Gallion, G. Debarge, and C. Chabran, "Locking bandwidth and relaxation oscillations of an injection-locked semiconductor laser," IEEE J. Quantum. Elect. **24**, 148-154 (1988).
18. O. Lidoyne, P. Gallion, C. Chabran, and G. Debarge, "Locking range, phase noise and power spectrum of an injection-locked semiconductor laser," IEE Proceedings J - Optoelectronics **137**, 147-154 (1990).
19. G. P. Agrawal, "Effect of gain and index nonlinearities on single-mode dynamics in semiconductor lasers," IEEE J. Quantum. Elect. **26**, 1901-1909 (1990).
20. K. Petermann, *Laser Diode Modulation and Noise* (Kluwer Academic Publishers, Dordrecht, 1988), p. 315.
21. G. Duan, P. Gallion, and G. P. Agrawal, "Effective nonlinear gain in semiconductor lasers," IEEE Photonics Technology Letters **4**, 218-220 (1992).
22. T. L. Koch and R. A. Linke, "Effect of nonlinear gain reduction on semiconductor laser wavelength chirping," Appl. Phys. Lett. **48**, 613-615 (1986).
23. T. Sasaki, Y. Yamamoto, and M. Koashi, "Practical quantum key distribution protocol without monitoring signal disturbance," Nature **509**, 475-478 (2014).
24. W. Engeler and M. Garfinkel, "Thermal characteristics of GaAs laser junctions under high power pulsed conditions," Solid State Electron. **8**, 585-604 (1965).
25. W. K. Nakwaski, "The thermal properties of a single-heterostructure laser diode supplied with short current pulses," Opt. Quant. Electron. **11**, 319-327 (1979).
26. W. Nakwaski, "Dynamical thermal properties of broad-contact double-heterostructure GaAs-(AIGa)As laser diodes," Opt. Quant. Electron. **15**, 313-324 (1983).
27. "Handbook series on semiconductor parameters," (World Scientific Publishing, Singapore, 1999).
28. S. Feng, X. Xie, W. Liu, C. Lu, Y. He, and G. Shen, "The analysis of thermal characteristics of the laser diode by transient thermal response method," in *1998 5th International Conference on Solid-State and Integrated Circuit Technology. Proceedings (Cat. No.98EX105)*, 1998), 649-652.
29. G. Oliveti, A. Piccirillo, and P. E. Bagnoli, "Analysis of laser diode thermal properties with spatial resolution by means of the TRAIT method," Microelectr. J. **28**, 293-300 (1997).
30. A. Piccirillo, G. Oliveti, M. Ciampa, and P. E. Bagnoli, "Complete characterisation of laser diode thermal circuit by voltage transient measurements," Electron. Lett. **29**, 318-320 (1993).
31. H. I. Abdelkader, H. H. Hausien, and J. D. Martin, "Temperature rise and thermal rise-time measurements of a semiconductor laser diode," Rev. Sci. Instrum. **63**, 2004-2007 (1992).
32. H. I. Abdelkader, "Electronic control of a semiconductor laser for an optical fibre packet LAN," (University of Bath, 1988).
33. Z. Fang, H. Cai, G. Chen, and R. Qu, *Single Frequency Semiconductor Lasers* (Springer Nature, Singapore, 2017).
34. A. McDaniel and A. Mahalov, "Stochastic Differential Equation Model for Spontaneous Emission and Carrier Noise in Semiconductor Lasers," IEEE J. Quantum. Elect. **54**, 1-6 (2018).
35. P. E. Kloeden and E. Platen, *Numerical Solution of Stochastic Differential Equations* (Springer, New York, 1995).
36. G. P. Agrawal and N. K. Dutta, *Semiconductor lasers* (Kluwer Academic Publishers, Dordrecht, 1993).